\documentclass[%
twocolumn,showpacs,amsmath,amssymb
]{revtex4}

\usepackage{xcolor}
\usepackage{graphicx}
\usepackage{dcolumn}
\usepackage{bm}

\begin{document}
	
	
	\title{Crumpling an elasto-plastic thin sphere}
	\author{Hung-Chieh Fan Chiang$^1$, Li-Jie Chiu$^1$, Hsin-Huei Li$^1$, Pai-Yi Hsiao$^2$, and Tzay-Ming Hong$^1$}
	\affiliation{$^1$Department of Physics, National Tsing Hua University, Hsinchu 30013, Taiwan, Republic of China\\
		$^2$Department of Engineering and System Science, National Tsing Hua University, Hsinchu 30013, Taiwan, Republic of China}

	\date{\today}
	
	\begin{abstract}
		
		The phenomenon of crumpling is common in our daily life and nature. It exhibits many  interesting properties, such as ultra-tough resistance to pressure with less than 30$\%$ of volume density, power-law relation for pressure vs density, and crackling noises with occurrence frequency vs intensity mimicking that of earthquakes. These results are mainly obtained by using flat thin sheets. But, in reality the majority of crumpled objects exhibit nonzero intrinsic curvatures. Notable examples are crushed aluminum cans, car wreckage, and cells move in and out of blood vessels. In this work we concentrate on establishing the fact that they behave very differently from flat sheets by employing both experiments and molecular dynamics simulations.

	\end{abstract}
	\maketitle

	\section{Introduction} 
	\color{black}
	
	Squeezing an object occurs quite often in our daily life, for example, crumpling a paper and crushing an aluminum can.
	Significant differences between the results of the above two examples have been observed: a wrinkled paper consists mainly of ridges and vertices whereas a squeezing ping pang ball shows mirror buckling on the surface.  
	Currently, main researches in the field are dedicated to crumpling flat sheets \cite{1,3,depth1,5}.
	Several interesting properties have been found.
	(1) The crumpled size $R$ of a sheet versus the applied pressure $P$ obeys a power-law relation  with an exponent independent of the thickness and the original size of the sheet \cite{self,7,8,10,11} but varying with the material made of it \cite{powerlaw,scaling}. 
	(2) As $R$ decreases, the crumpled sheet forms ordered layering domains inside the object\cite{spontaneous} and
	the change of the internal structure causes a break-down of the power law of $R$ vs $P$, which is replaced by a regime where different data sets can be mapped onto a master curve \cite{scaling}. And (3) energetics study for the ratio of the stretching and the bending energies $E_b/E_s$ on each ridge shows
	the power law relationship between total energy $E(\bar\ell)$ and the average ridge length $\bar\ell$\cite{Witten,change}.
	In this work we employ  experiments and molecular dynamics (MD)\cite{LAMMPS} simulations to study the properties of 3-D crumpling of a spherical shell and compare its properties with those obtained from squeezing a flat sheet.

	\begin{figure}
		\centering
		\includegraphics[width=8.5cm]{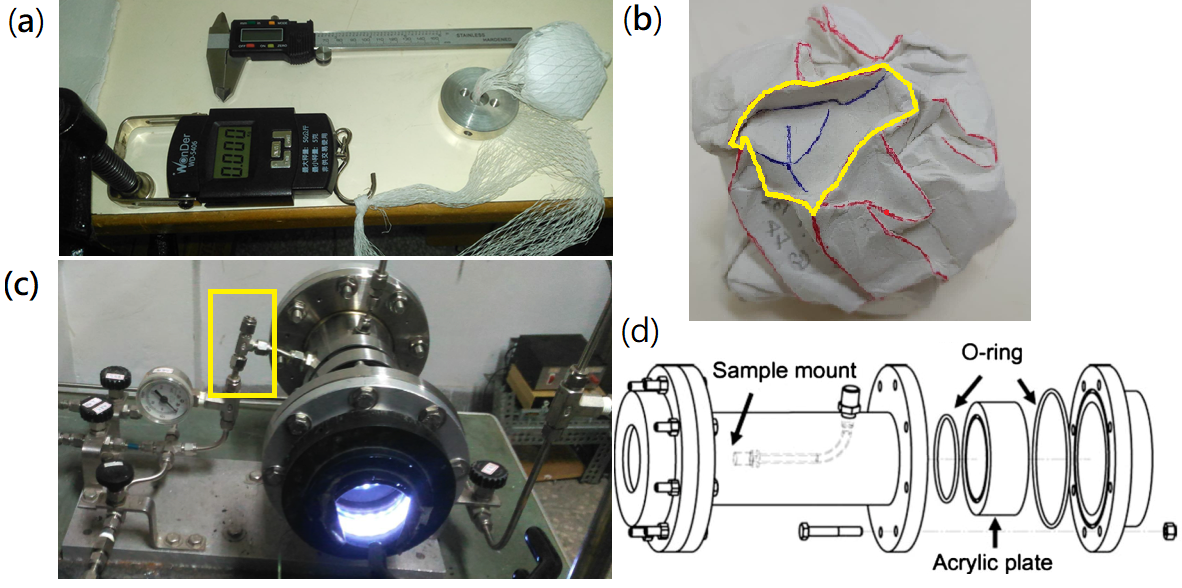}
		\caption{(color online) The experimental set-up (a) the net connect to the force gauge and pass through the a steel ring. Crumpled sample (b) after using net shows enclosed ridge, which is rare for thin sheets. High pressure experimental set-up (c) yellow box shows the steel tube connected the sample to outside and (d) schematic diagram. }
	\end{figure}
	
	\color{blue}
	For our experiment the spherical shells are home-made with paper pulp of 80 gsm. We use net to pack the sample and connect with a force gauge as shown in Fig.1(a). When it is hard to pull the net through the ring, we keep the same pressure of the sample, as Fig.1(b), and put it into high pressure chamber. Schematic plots of our pressure chamber are shown in Fig.1(c, d)\cite{scaling}. By connecting the wrapped sample by  a steel tube to outside, an ambient pressure difference is created that crumples the sample when high density nitrogen gas flows into the chamber. The steel-made chamber can sustain pressure up to 120 PSI with leakage flow rate of about 71.7 ml/h.

	We first study experimentally the variation of the size $R$ of the crumpled spherical shells with the compressing pressure $P$. To compare the data under different condition, we plot the dimensionless pressures $P/P_0$ versus the size ratio $R/R_0$ in Fig.2(a), where $P_0, t$ and $R_0$ denotes the first detected compressing pressure thickness and initial size of sample. As we keep the same compressing pressure on sample, we found the size of sample reduce suddenly and miss the data during this process. However, we can separate Fig.2(a) in three regime: (1) low-pressure regime that can be compressed by using net, (2) unknown regime, and (3) high-pressure regime that can only be compressed by high pressure chamber. These three regime depends on compressing pressure, which is similar to compressed flat sheet case\cite{scaling}. If we consider these three regime as same mechanical properties, we should observed power-law in regime 1, and master curve in regime 3. On one hand, we found a power-law relation $P/P_0\sim(R/R_0)^{-\alpha}$ with exponent $\alpha \sim 4.59$. On the other hand, we fail to mapped data onto a master curve in regime 3. In this case, we try to use MD simulations to figure out the detail of the compressed process.
	
	\color{black}	
	\section{molecular dynamics simulation}
	
	For our simulation, Weeks-Chandler-Anderson potential \cite{WCA} is used to model the excluded volume of each lattice point \cite{self,plastic} which defines length unit ($\sigma$) and energy unit ($\epsilon$).
	We follow the convention of  previous MD simulations on flat thin sheets by adopting a hexagonal lattice with the mean spacing $L_0=1(\sigma)$ \cite{self,plastic,change, LAMMPS}. 
	The spherical shell is constructed by arranging particles periodically along each longitude. 
	As discussed in Ref.\cite{defect,defect0,defect2}, it is generally not possible to arrange the lattice points regularly  on a spherical surface and, therefore, some defects can occur at lattice points connected to five or seven neighbors.  
	We have checked that the percentage of these defects is less than $2\%$ which safeguards the credibility of the simulation results. 
	The spherical shell is enclosed inside a cavity by an impenetrable wall.
	Crumpling the shell is effectuated by reducing the radius $R(\sigma)$ of the cavity. 
	The effective thickness $0.73\leq t(\sigma)\leq 1.46$ of the shell is given by $k_b/k_s = 3 t^2 /32$ \cite{change} where $k_b (\epsilon)$ is the bending modulus and $10^4 \leq k_s (\epsilon/\sigma^2) \leq 2\times 10^4$ the stretching modulus.
	Different materials are simulated by varying $k_s$ but keeping the ratio $t$ constant.
	The work stored in the deformed shell comprises two forms: the stretching energy $E_s(\epsilon)=k_s(L-L_0)^2/2 $ and bending energy $E_b(\epsilon)=k_b(\theta-\theta_0)^2/2 $ where 
	$L(\sigma)$ is the length between adjacent points and 
	$\theta$ is the angle spanned by three consecutive beads along a direction and 
	$\theta_0$ is equilibrium angle.
	Plasticity is included by halving the magnitude of $k_b$  beyond a yield angle $|\theta -\theta_0 |$ of $10^{\circ}$. 
	In our simulation, we set compressive rate $\tau_v=-0.001(\sigma/{\rm time})$.
	
	\begin{figure}
		\centering
		\includegraphics[width=8cm]{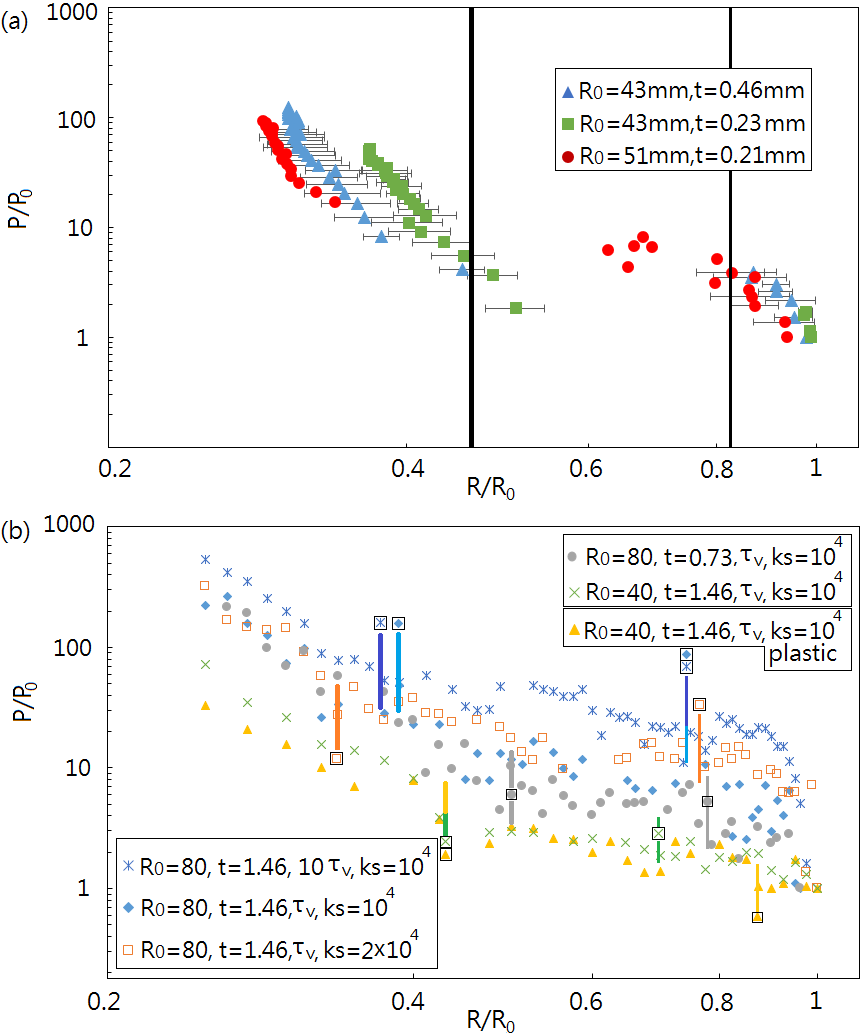}
		\caption{(color online) Full-log relation between $P/P_0$ and $R/R_0$ for a compressed spherical shell is found to be insensitive to $k_s$, $R_0$, and thickness. Thin vertical lines signalize the boundary between regime 1 and 2, while thick lines mark the transition to regime 3. The power-law regime at low density can be fit to $P/P_0 \sim (R/R_0)^{4.59}$ for experimental data in (a)  and $P/P_0 \sim (R/R_0)^{4.83}$ for simulation data in (b).}
	\end{figure}

	\section{density versus  pressure}
	
	\color{blue}
	Now we check the relation of $P/P_0$ vs. $R/R_0$ and separate it into three regime as shown in Fig.2(b) in the same way as experiment before. To compare all variable of sample, we change $R_0, t, \tau_v, k_s$, and plastic or not. We first check the power-law relation in regime 1 and get $P/P_0\sim(R/R_0)^{-\alpha}$ with exponent $\alpha = 4.74 \sim 4.83$. In regime 2, we found $P/P_0$ behave nearly a plat, which give a reason why we keep same compressing pressure on sample but still miss those data. Although we already rescale the axes, data still can not mapped onto a master curve. All these qualitative properties are same as experiment result, we can look in detail of simulation result.
	
	For regime 1, we compare the $\alpha$ value of spherical shell $4.59\sim 4.83$ to the flat sheet case $4\sim 4.5$. This causes the resistance pressure by a crumpled spherical shell to shoot up faster than a flat sheet.
	The power-law break down in regime 2 where the initially smooth rims of indentations start to touch each other and deform into polygons.
	Since we already know the data cannot be mapped to a master curve, the structure of the spherical shell in regime 3, as shown in Fig.3(a), is similar to flat sheet. The only reason that master curve disappear is that the curvature is a local property that can not be erased by compact packing.
	
	The overall view of compressed spherical shell can observe that the compressing pressure is about one and half times stronger than the flat sheet of a similar size \cite{scaling}. The boundaries of three regime, the thin and thick vertical line, happen earlier than the flat sheet.

	\color{black}
	\section{energetic scaling}
	Base on the simulations result, we can discuss the energetic scaling of compressed spherical shell. 
	The relation between $E_b/E_s$ and $R/R_0$ is shown in Fig.3(b) which also reveals a change of trend whenever transitions between different regimes occur. In more details, (1)  there is generally a dip in the regime 1, consistent with Ref.\cite{simusphere}, when the rim of indentation is smooth the stretching energy increases then starts to decrease when rim become polygons.
	Exception occurs for (i) a slow compression rate that favors the creation of only a few indentations that grow in size $\ell$. The base area $\propto \ell^2$ is rich in $E_b$ and quickly trumps the perimeter$\propto\ell$ where $E_s$ resides; and (ii) an ultra-thin sphere (see grey circles) that generates many  small indentations whose base area is too small to  dominate its perimeter.
	(2) Upon entering the mixed regime, $E_b$ increases faster than $E_s$. This can be understood by the fact that the number of indentations already saturate by now, and further crumpling can only prompt them to grow in size and thus give $E_b$ the edge - same argument as in (1), and (3) when the crumpled spherical shell crosses into the compact regime,  $E_s$ suddenly becomes important and causes $E_b/E_s$ to drop. This is the same physics behind why paper is hard to fold by more than  8 times. 
	Both regime boundaries happen earlier with thickness, compression rate, and the inverse of size and hardness, but 2nd boundary is less sensitive. Furthermore, 1st boundary and 2nd boundary are roughly one third and one half of those for flat sheets\cite{scaling}. All these properties are new, since the ratio of $E_b/E_s$ varies with thickness, size, and compression rate in contrast to being independent for flat sheets\cite{change}.
	
	\begin{figure}
		\centering
		\includegraphics[width=8.5cm]{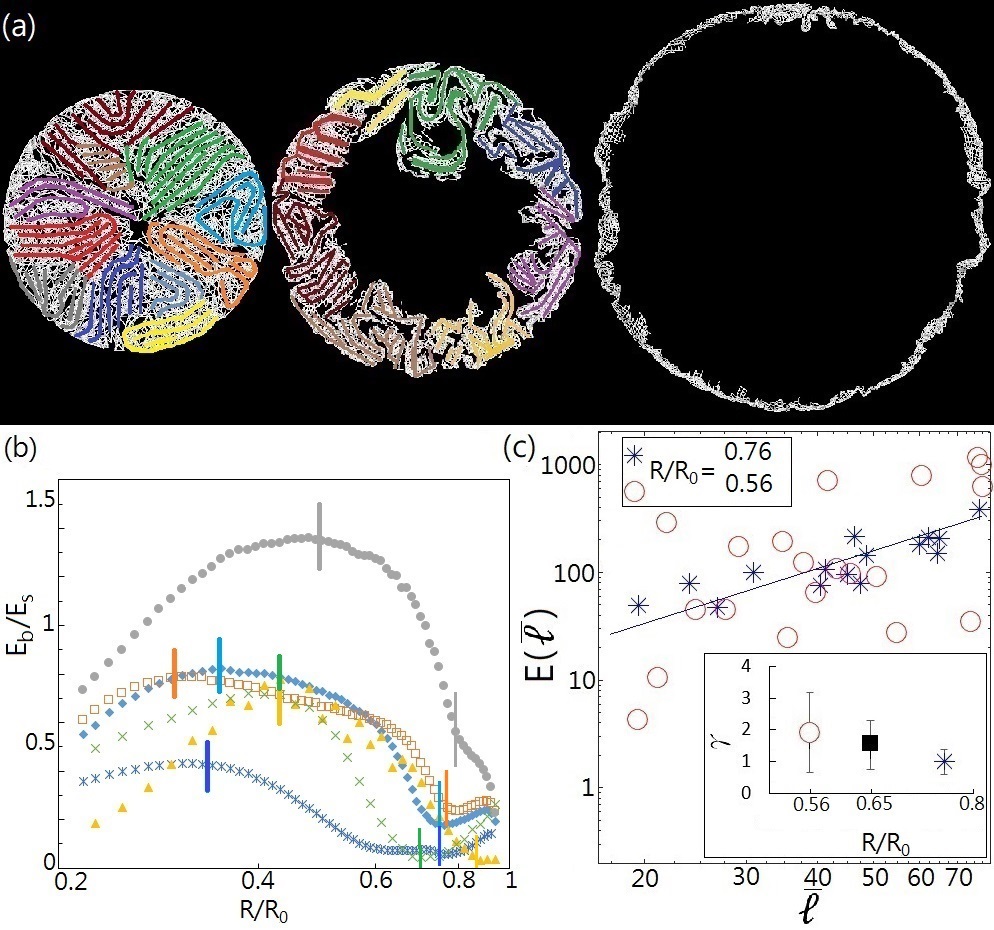}
		\caption{(color online) Cross sections at the big circle are shown in (a) for different stages of a  crumpled thin sphere in simulations. As $R/R_0$ decreases from right to left, curved surfaces or the nonzero intrinsic curvature remain a vital component of the internal structure. (b) Semi-log plot of $E_b/E_s$ vs $R/R_0$ for a crumpled spherical shell with different $k_s$, $R_0$, thickness and compression rate, these labels are same as Fig.2. Full-log plot (c) checks the validity of energetic scaling of  $E(\bar\ell )\propto \bar\ell^\gamma$ and concludes that $\gamma$ increases from 1 to 1.94 as $R/R_0$ decreases from 0.8 to 0.5 for $R_0=80$, $t=1.46$, $\tau_v$, and $k_s=10000$.}
	\end{figure}

	Both mechanical property and the ratio of bending and stretching energy variance are whole spherical shell properties, the local deformation are still unknown yet. We define $\ell$ the rim of the indentation represent its size and found the relation between its stored energy $E(\ell)$ and it size as shown in Fig.3(c). However, the indentations have different size so we choose to use mean size $\bar\ell$ for instead then we can write $E(\bar\ell)\propto \bar\ell^\gamma$.
	According to Fig.3(c) the initial value of $\gamma =1$ and break down when $R/R_0$ decrease. Now we have two things should discuss: (1) is $\gamma =1$ reasonable, and (2) what happen to the spherical shell and break down the relation? (1) Based on previous studies, we already know $\gamma=1/3$ for flat sheets \cite{Witten,change} when deformations are scant. This is intimately related to the shape of deformation. Each ridge on a flat sheet contains two vertices at both ends. When the ridge length is doubled, no extra work is required to generate further vertices so the  work is less than doubled. But the rim of an indentation is closed and contains no vertex. Therefore, $\gamma =1$ is reasonable.
	(2) When $R/R_0$ equal to 0.76, we found that is around the boundary of regime 1 and 2, namely, the rim start to deform into polygon. Further compressed the spherical shell, most rim touch each other and rim-rim interaction increase then break down the energetic scaling, the growth of error bar imply the same reason.

	\begin{figure}
		\centering
		\includegraphics[width=8cm]{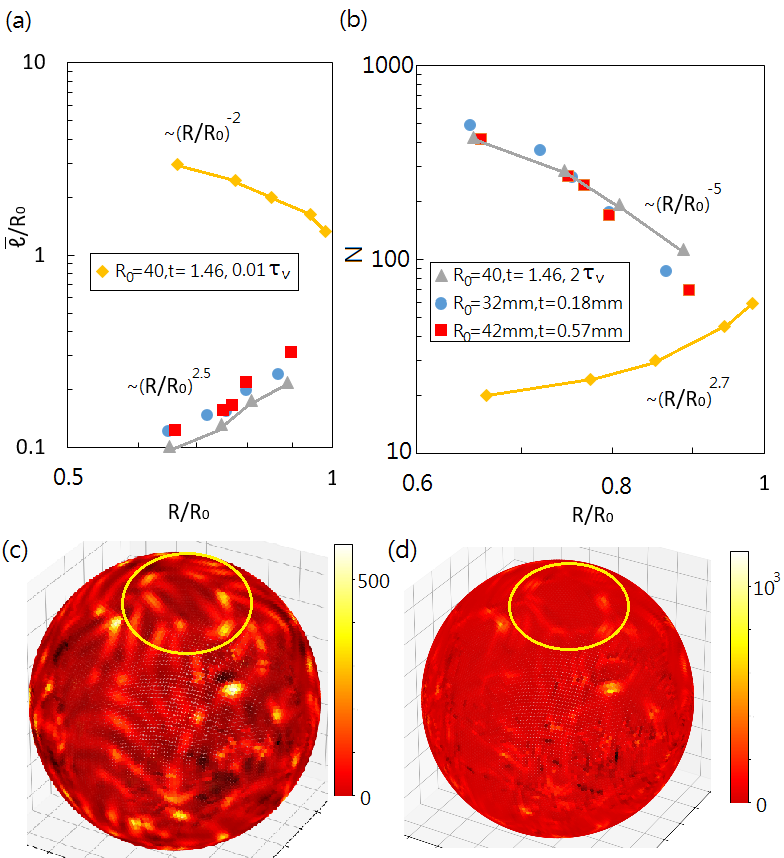}
		\caption{(color online) We show both MD simulation (yellow diamond and grey triangle) and experiment result (red square and blue circle) of (a) $\bar\ell$ vs. $R$ and (b) $N$ vs. $R_0/R$. Note that these MD simulation result have same material property $k_s=10000$ but in different compression rate then give different tends. Contour maps for bending and stretching energies from MD simulations are plotted respectively in (c) and (d) for $R_0=40, t=1.46, \tau_v$, and $k_s=10000$ at density $R/R_0=0.75$. The yellow circle highlights the different distribution of $E_b$ and $E_s$ at the same indentation where the former concentrates mainly on the base area, while the latter at the rim.}
	\end{figure}
	
	\section{theoretical analysis}
	
	Let's now move on to studying the evolution of $N$ and $\bar\ell$ of rims. Similar to flat sheets\cite{change}, we found that $N{\bar\ell}^2$ is roughly a constant when $\tau_v$ that favors creating enough indentations to cover the whole surface, while $\bar\ell$ and  $N$ vary in a power-law fashion with $R$ in Fig.4(a, b). Instead of 1 and -2 for  flat sheets, their corresponding exponents equal 2.5 and -5, respectively. On the other hand, when compressed rate $<0.1\tau_v$, the indentations exist only in partial area of the sphere  in the initial stage and $N\propto R^{2.7}$ and $\bar\ell\propto R^{-2}$.
	
	By use of the relations $N\sim (R_0 /R)^5$ and $\bar\ell/R_0\sim (R/R_0 )^{2.5}$, the total storage energy $E  \sim N{\bar \ell}^\gamma$ can  be obtained as $E\sim R_0^{5-\frac{3\gamma}{2}}R^{\frac{5\gamma}{2}-5}$. Differentiating $E$ with respect to $R$ gives the crumpling force $F$. Further dividing $F$ by the surface area $4\pi R^2$ then renders the pressure $P\sim R_0^{\frac{3\gamma}{2}-5}R^{\frac{5\gamma}{2}-8}$. This  prediction is consistent with Fig.2 as $\gamma$ increases from 1 to 2 according to Fig.3(c). This evolution of $\gamma$ again can be understood by realizing there are two contributors to  $E(\bar\ell)$, i.e., $E_{\rm rim}\propto\bar\ell$ and $E_{\rm base}\propto\bar\ell^2$. The former dominates when $\bar\ell$ is small that suits the regime 1, while the latter is more important when $\bar\ell$ is large that applies to the regime 2. This completes our heuristic argument for the results in Fig.3(c) and of $P/P_0$ vs $R/R_0$ in Fig.2.
	
	According to the results of Fig.4(a), we are in a better position to discuss the origin of a dip in Fig.3(b). Separating the contribution to $E(\bar\ell)$ from the base and rim tells us $E_b/E_s\sim E_{\rm base}/E_{\rm rim}\sim \bar\ell\sim R^{2.5}$. Easy differentiation gives $\frac{d(E_b/E_s)}{d(R/R_0)}\sim R^{1.5}$ give a positive coefficient and explains the dip. On the other hand, Fig.4(b) informs us that how $\bar\ell$ and $N$ change with $R/R_0$ in slow compression rates is totally different from that in fast rates. This provides evidence for the argument we discussed in $E_b/E_s$ vs. $R/R_0$ (1)(i) part.

	\section{conclusion}
	In this paper we combine experiments, MD simulation, and theory to study the mechanical, energetic, and statistical properties for a crumpled thin sphere. Our results turn out to be very different from and complement the previous knowledge obtained from studying flat sheets. Considering that the majority of deformed objects in our daily life exhibit nonzero Gaussian curvatures, such as car wreckage and crushed aluminum cans, our findings can be very critical and useful. Our main conclusions are (a) the dependence of $E_b/E_s$ on $R/R_0$ is not monotonic for spheres, as opposed to decreasing from an initial value of 5 to around 2 for flat sheets, (b) the exponent $\gamma$ in $E(\bar\ell)\sim {\bar\ell}^{\gamma}$ increases from 1 to 2 for spheres, instead of 
	from 1/3 to 1 for flat sheets, (c) three different regimes can be defined in $P/P_0$-$R/R_0$ relation both flat sheets and spheres, but the transitions happen earlier, i.e, at a smaller critical $R/R_0$ and there is no master curve in spheres. This results in a shrinkage of valid range for the power law in spheres,  (d) the evolution of $N$ and $\bar\ell$ with $R$ obey the power-law relation, but the exponents equal -5 (2) and 2.5 (-2.7) for fast (slow) compression rates, as opposed to -2 and 1 for flat sheets, (e) all the properties of crumpled spheres are sensitive to compression rate, thickness and size of material, instead of being independent in the case of flat sheets.

\end{document}